# Causal loop and Stock-Flow Modeling of Signal Transduction Pathways

**Sadegh Sulaimany[1,*], Gholamreza Bidkhori[2], Sarbaz H. A. Khoshnaw [3]**

**[1]Department of Computer Engineering, University of Kurdistan, Sanandaj, Iran**

**[2]Royal Institute of Technology (KTH) - Science for Life Laboratory, Seweden**

**[3] Department of Mathematic, University of Raparin, Ranya, Kurdistan Region, Iraq**

***Corresponding Author**
Sadegh Sulaimany, Ph. D.
Department of Computer Engineering
University of Kurdistan, Sanandaj, Iran
E-mail: S.Sulaimany@UoK.ac.ir
WWW: http://Bioinformation.ir
Tel: +98-87-33611410 Fax: +98-87-33668513
ORCID: 0000-0002-4618-0428

## Abstract

System dynamics is a popular approach in many fields of science and technology, but it has not been investigated for cell signaling pathways yet. It is a well formulated methodology used to analyze the components of a system considering the cause-effect relationships. The two main components of system dynamics modeling, Causal Loop and Stocks-Flow diagrams, make possible to model the dynamics of the system with the ability to analyze it both qualitatively and quantitatively.

In this paper, after introducing the system dynamics modeling approach, and giving a simple example from its usage for Michaelis Menton reactions, a three-step process is proposed for signal transduction modeling. Then in a complete example, it is applied to a case study in cell signaling pathways, namely the RKIP Influence on the ERK signaling pathway and the results is compared with an integrative modeling approach based on Petri net and ODEs.

Computational simulations show the success of the system dynamic in easier and effective modeling of the cell signaling pathways, in addition to its diverse options for understanding and testing the pathway quantitatively and quantitatively at the same time in relation to other methods. More intestinally, the suggested approach here helps one to identify all biochemical reaction paths and loops for complex cell signaling pathways. It will be a good step forward to define dominant systems for such complex cell signaling pathways.

## 1. Introduction

Systems biology involves the network aspects of biological networks. Its scope can vary from a single cell reaction to whole organism simulation. Mathematical modeling plays important role in modeling and converting biological interactions into quantitative data for related analysis. Such models try to simulate the system as much as similar to real one. If there exist a model that can simulate the biological system in both quantitative and qualitative views, it will present a more precise description from the reality.

Many efforts have so far been made in order to reduce the complexity of cell signaling pathway analysis by creating or proposing new suitable methods and models. Summary of different computational approaches for the modelling of cellular signaling is as follows (Pablo, Pérez, & Lagúnez-otero, 2000):

1. Boolean Networks: This logic based method demonstrates signal transduction networks in a simple manner without taking into account the detailed biochemistry of the interactions. Each interaction works as an OR, AND or others logical gates and composing the different interaction with various wirings, depicts and simulates the network.
2. Expert systems (Rule based): The interactions of the cell signaling network feed to the model in case of some rules with predefined formats for knowledge representation. After building the knowledge base of the model, an inference engine, makes the inferences and relates and runs some rules, and models the biochemical reactions according to user needs and requests.
3. Petri nets: This is a special graph type with extended capabilities to model complicated relations of the signal transduction networks. Rather than graphical representation of the network, it gives temporal information of signal propagation dynamics. However it is relatively complicated to learn and implement.

4. Agent based modeling: For the quantitative simulation of signaling pathways, protein messengers can be imagined as agents and reactions can be proposed as interaction between them. This type of simulation considers the effects of the agents on the cell system as a whole also. Each agent obeys simple rules and contacts with other agents in order to simulate the collective behavior of the cell.
5. Artificial Neural Network: This type of modeling, simulates the cell as molecular computer that get the input stimulus and process them and gives the output to another adjacent cells and the environment. ANN is not sensible for end user and does the calculations in a special manner such that only machine learning specialists can configure and run the model. It needs several real samples to be trained.
6. ODEs: It is the basic of many modeling techniques. The dynamics of a signaling pathway is often modeled as a system of nonlinear ordinary differential equations (ODEs) (Koh, Hsu, & Thiagarajan, 2010), using mass action kinetics, which can be integrated to determine the concentration of species over time. Direct mathematical modeling is not easy to understand for non-mathematician in biology and chemistry science.

There are also methods from other fields of science which has not been investigated for cell signaling pathways yet and some of them may be worth applying. System dynamics is a well formulated methodology for analyzing the components of a system including cause-effect relationships and their underlying mathematics and logic, time delays, and feedback loops. It began in the business and industrial world, but is now used in many other disciplines (Nuhoğlu & Nuhoğlu, 2007).

Presenting biological processes for mathematical modelling is normally based on the idea of classical theory of chemical kinetics that is called mass action law. The assumption is that a signaling pathway model consists of three components. The first vector of components is species, $S = \{S_1, S_2, \ldots, S_m\}$. This is sometimes called the set of states or types. The second vector is reaction rates, $V = \{v_1, v_2, \ldots, v_n\}$. The third set of components is parameters (kinetic constants), $K = \{k_1, k_2, \ldots, k_n\}$. For an elementary chemical reaction with m species that is given below

$$\sum_{j=1}^{m} a_{ij} S_j \underset{k_i^b}{\overset{k_i^f}{\rightleftharpoons}} \sum_{j=1}^{m} b_{ij} S_j, \quad i = 1, 2, \ldots, n. \tag{1}$$

Where $a_{ij}$ and $b_{ij}$ are non-negative stoichiometric coefficients. , We use the standard mass action law to define the rate of reactions. The reaction rates are given below

$$v_i = k_i^f \prod_{j=1}^{m} S_j^{a_{ij}}(t) - k_i^b \prod_{j=1}^{m} S_j^{b_{ij}}(t), \quad i = 1, 2, \ldots, n, \tag{2}$$

Where $k_{if}$ and $k_{ib}$ are reaction rate coefficients. The stoichiometric vectors for equation (1) take the form $h_{ij} = b_{ij} - a_{ij}$ . The differential equations can be used to describe the dynamics of chemical reactions. Such equations are given:

$$\frac{dS}{dt} = \sum_{i=1}^{n} h_i v_i. \tag{3}$$

The aim of this study is to show the feasibility of the application of system dynamics approach to the study of cell signaling pathways. The first part of the study explains the system dynamics concepts, component and customized modeling process. In the second part, applications of system dynamics in cell signaling pathway modeling and analysis are studied. One sample pathway, RKIP Influence on the ERK Signaling will be modeled and results will be compared to the most popular modeling technique, ODE. Finally, the benefits of using system dynamics to model cell signaling pathways are discussed. More interestingly, the applications of Vensim for cell signaling pathways are studied. This is an important computational tool to identify signaling pathways and loops for complex biochemical reaction networks.

## 2. System Dynamics Modeling

The conception of development of system dynamics took place during the late 1950's at the Massachusetts Institute of Technology under Forrester, with first applications in management (Wolstenholme, 1983). System dynamics modeling, is a well-established methodology especially for studying and managing complex feedback systems, and has been used to address a wide variety of science and engineering studies including (Conference, 2006; Wolstenholme, 1983): management, economy, urban, physics, control engineering, mechanics, chemical engineering and etc.

Important biological networks can be one of the following (Gilbert & Heiner, 2006; Sulaimany, Khansari, & Masoudi-Nejad, 2018): Metabolic Networks, Gene Regulatory Networks, Signal Transduction Networks and Protein-Protein Interaction Networks (PPIN). Metabolic networks describe the basic biochemistry in a cell. The metabolic process can be characterized as a set of biochemical transformations, each of which involves the consumption of one or more metabolites and the production of one or more metabolites. System dynamics has been applied in biochemistry and cell biology mainly in manual of Berkeley Madonna system dynamic modeling software and also some educational materials on the software’s website (Macey & Oster, 2006). There are also some papers using system dynamics for metabolic network modeling (Al-Akwaa, 2006; Demirezen & Soc, 2009), but it has not yet been applied specifically for signal transduction pathway modeling and analysis such systems.

To grasp the essence of system dynamics it is better to get a sense of what the system dynamic does, especially in the context of comparing different modeling methods? Consider a modeling of a system that can simultaneously depict its characteristics from two different views; qualitative and quantitative. In qualitative view one can understand the different relationships, such as feed forward and feedback between the elements of the system visually and get a quick understanding. On the other hand, in quantitative view, it can be found the exact effects of the elements on each other and identifying a deeper perception of the system and its elements and their causal relationships numerically. One of the key advantages of the system dynamics approach is to elegantly and clearly identify feedback loops, in addition of other relations, and include them into the model (Borshchev et al., 2004). Such property is favorable to modeling signal transduction systems.

### 2.1. Major Components in System Dynamics Models

Two major components of a system dynamics model are the Causal Loop and Stock-Flow Diagrams. While the former has a qualitative nature the later has the quantitative one, final model of the system is a combination of both. We will explain them briefly here:

1. Causal Loop

The concept of information flow, or feedback loops, is the heart of the system dynamics approach. It helps to capture a hypothesis about the causes of dynamics quickly, eliciting and capturing the mental models, communicating the important feedbacks believed to be responsible for the problem by connecting system variables with arrows in a proper manner. Loops can be positive feedback, strengthen the signal, or negative feedback, decreasing the effect or inhibit it. A sample causal loop is shown below (Fig. 1).

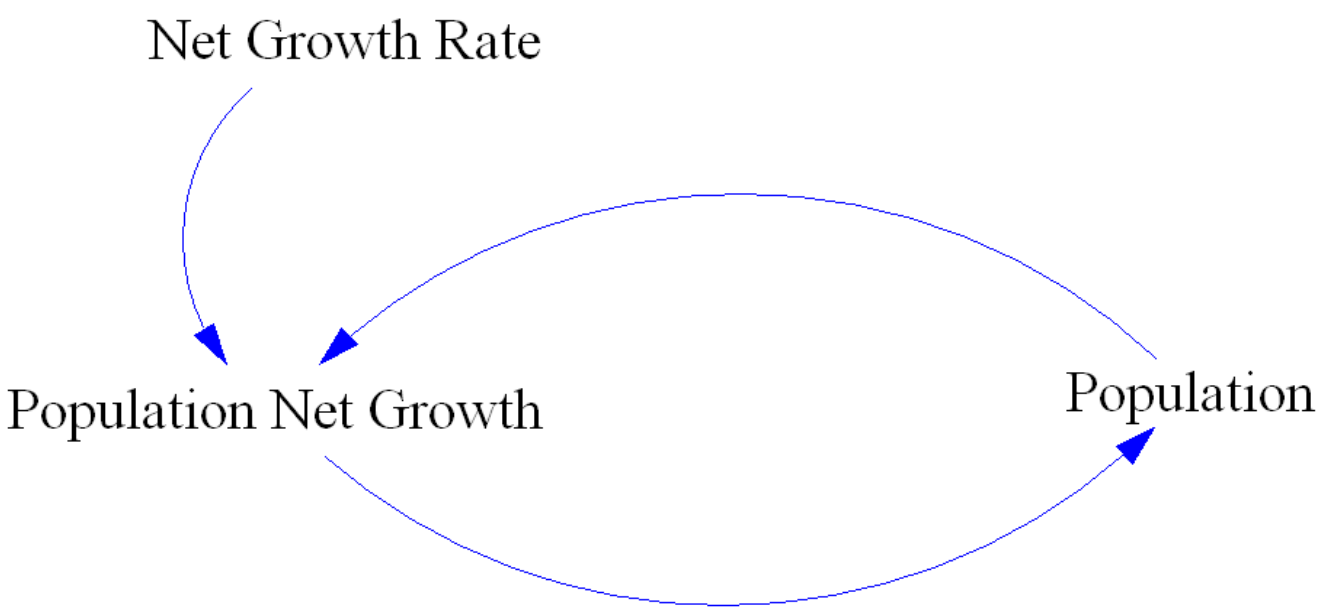

**Fig. 1.** A simple example of causal loop diagram for population; Population Net Growth has effect on Population, while it is affected by both Population and Net Growth Rate

2. Stock-Flow

The loop concept underlying feedback and circular causality by itself is not sufficient. We also need to know and collect detailed data of system behavior and simulate steady state or stability conditions of the system. Stocks (levels) and the flows (rates) that affect these stocks are essential components of the system structure in system dynamics. Stocks are the accumulations or state variables, which are the memory of a dynamic system and are the reasons of its disequilibrium and dynamic behavior. In the following picture (Fig. 2), Stock is a sample for **Stock** or level and Inflow and Outflow are the **Flow** or rate.

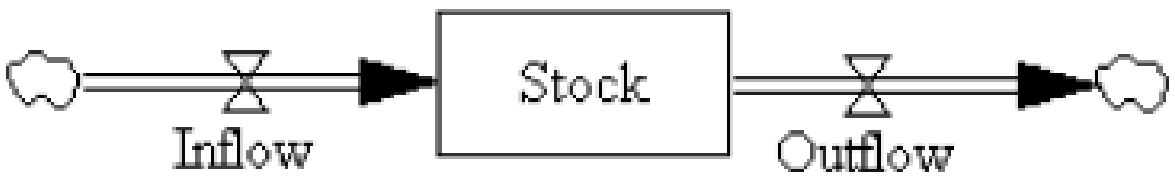

**Fig. 2.** An Example of a simple Stock-Flow diagram for an inventory

Users and researchers only need to learn to draw the diagrams and set the parameters in the associate software. Each diagram has an, automatically generated equivalent source code representing the model parameters and formulas enabling the user to understand and probably regenerate the same model again. In the simple model presented in Fig. 2, the associated code, abstracted without details, generating by modeling software is as follows:

*(1) FINAL TIME = 100*

*Units: Second*

*The final time for the simulation.*

*(2) Inflow = A FUNCTION OF( )*

*Units: **undefined***

*(3) INITIAL TIME = 0*

*Units: Second*

*The initial time for the simulation.*

*(4) Outflow = A FUNCTION OF( )*

*Units: **undefined***

*(5) Stock = A FUNCTION OF( Inflow,-Outflow)*

*Units: **undefined***

*(6) TIME STEP = 1*

*Units: Second [0,?]*

*The time step for the simulation.*

Completing stock-flow diagrams with causal loop diagrams will lead to a final simulation model, the Dynamic Model. This can be used to fully analyze the model. Fig. 3 is a sample dynamic model of a reversible first order reaction and the followings is the associated model parameters and formulas.

$$A \underset{kr}{\overset{kf}{\leftrightarrow}} B$$

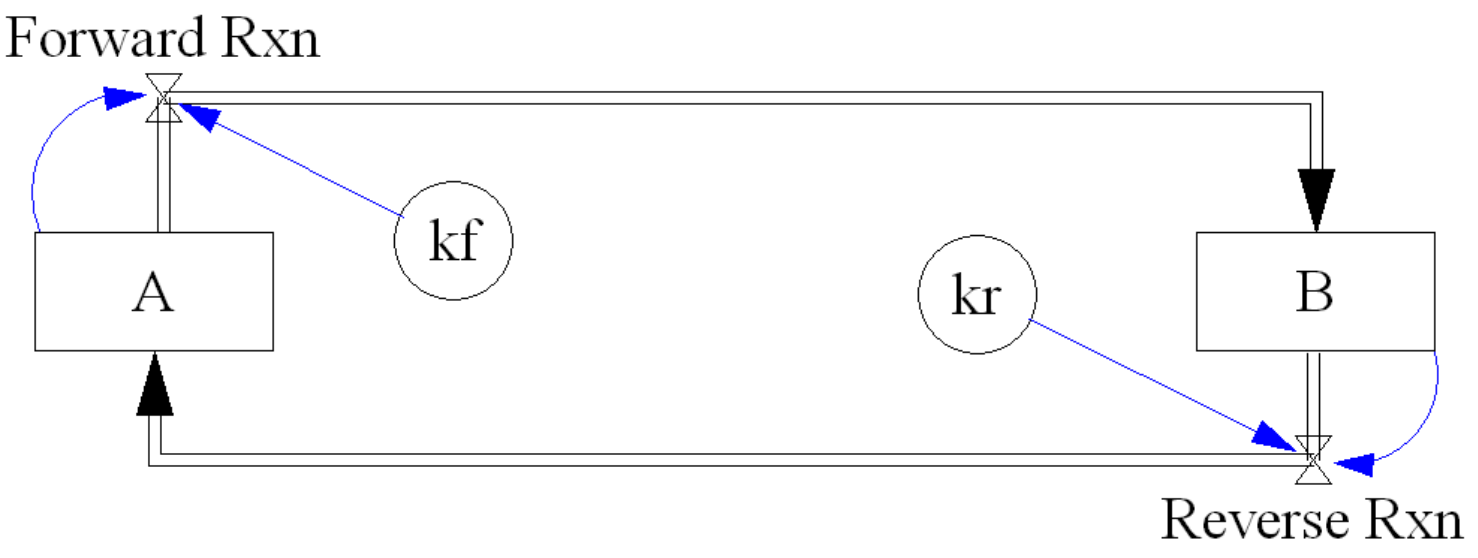

**Fig. 3.** System Dynamic Model of a reversible first order reaction

Simulation code is self demonstrating and easy to understand. The only expression which needs to be clarified is INTEG. INTEG represents the integral of a variable during the simulation time starting from an initial value.

```
(01)    A= INTEG ( Reverse Rxn-Forward Rxn, 1000)
(02)    B= INTEG ( Forward Rxn-Reverse Rxn, 0)
(03)    FINAL TIME  = 10
        Units: Second
        The final time for the simulation.
(04)    Forward Rxn= A*kf
(05)    INITIAL TIME  = 0
        Units: Second
        The initial time for the simulation.
(06)    kf= 2
(07)    kr= 1
(08)    Reverse Rxn=  B*kr
(09)    TIME STEP  = 0.0625
        Units: Second [0,?]
        The time step for the simulation.
```

Extending the case to cover more complex examples in signal transduction modeling is easy. Such models are created using special building blocks iteratively. Therefore, we show how to create the building block of almost all enzyme reactions, Michaelis Menton reactions, in cell signaling pathways to illustrate the extendibility of the approach. More complicated examples are composed of using and connecting this building block iteratively (Cho, Shin, Lee, & Wolkenhauer, 2003). The following example is a popular model of enzyme kinetics.

$$E + S \underset{k2}{\overset{k1}{\leftrightarrow}} ES \xrightarrow{k3} E + P \quad (4)$$

In equation 4, E is the concentration of an enzyme that combines with a substrate S to form an enzyme-substrate complex ES with a rate constant $k_1$. The complex ES holds two possible outcomes in the next step. It can be disassociated into E and S with a rate constant $k_2$, or it can further proceed to form a product P with a rate constant $k_3$. It is assumed that none of the products reverts to the initial substrate. The model equations (ODE) are given based on mass action law, simulation software will automatically calculate the formulas behind the model:

$$\begin{aligned}
dE(t)/dt &= -k_1.E(t).S(t) + (k_2 + k_3).ES(t) \\
dS(t)/dt &= -k_1.E(t).S(t) + k_2.ES(t) \\
dES(t)/dt &= k_1.E(t).S(t) - (k_2 + k_3).ES(t) \\
dP(t)/dt &= k_3.ES(t)
\end{aligned} \quad (5)$$

Our proposed system dynamic model and its output diagram with some arbitrary values is as follows. It is obvious that simulation details are depending on the tool and the user.

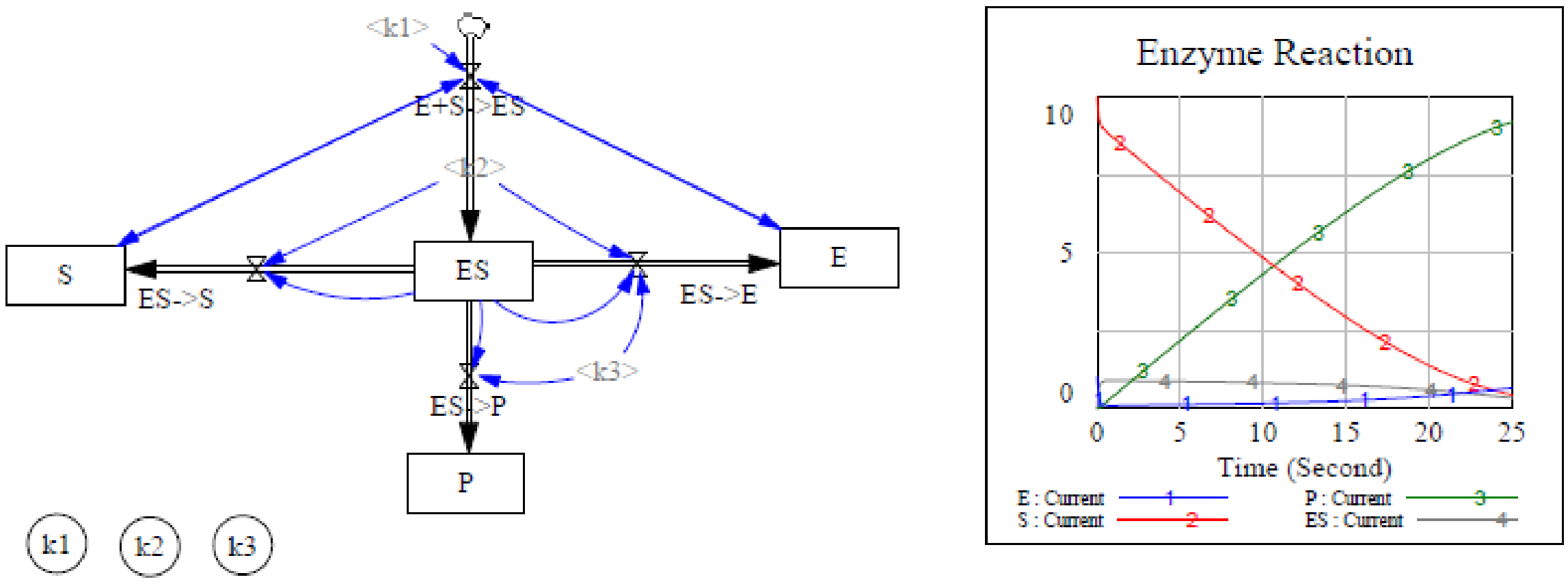

**Fig. 4.** Reaction profile in system dynamics of basic enzyme kinetics

*(01)* *E= INTEG ( "ES->E"-"E+S->ES", 1)*

*(02)* *"E+S->ES"= k1*E*S*

*(03)* *ES= INTEG ("E+S->ES"-"ES->E"*2+"ES->P"+"ES->S", 0)*

*(04)* *"ES->E"= (k2+k3)*ES*

*(05)* *"ES->P"= ES*k3*

*(06)* *"ES->S"= ES*k2*

*(07)* *FINAL TIME = 25*

*Units: Second*

*The final time for the simulation.*

*(08)* *INITIAL TIME = 0*

*Units: Second*

*The initial time for the simulation.*

*(09)* *k1= 1*

*(10)* *k2= 0.5*

*(11)* *k3= 0.5*

*(12)* *P= INTEG ( "ES->P", 0)*

*(13)* *S= INTEG ( "ES->S"-"E+S->ES", 10)*

*(14)* *TIME STEP = 0.125*

*Units: Second [0,?]*

*The time step for the simulation.*

## 2.2. Modeling Steps

There are several approaches to develop system dynamics models based on the aforementioned components (Pejić-Bach & Organizational, 2007). Some of them first use the causal loop to start the modeling and others first use stock-flow. There are some situations which researcher do not have enough information about the problem and needs to refine the structure of the solution several times when new information is revealed during modeling. These situations are more time-consuming for modeling and have more steps to simulate. But for cell signaling pathways, we are often equipped with the preliminary experimental data and information first and analyze the pathway computationally next.

It is better to start with an approach based on the causal loop diagram which seems better fitted for cell signaling modeling regarding to the authors experience. For the first step, we propose building qualitative model with the system dynamics software using causal loop diagrams that are suitable for explaining the overall model structure. Then one can directly develop a dynamic model consisting of casual loops and stock-flow diagrams, quantitative model. That is because of the type of the problem we are facing with, having reaction equations or related differential equations plus kinetics rates and initial concentration before; it is straightforward to draw the final dynamic model. There is a big advantage here that makes system dynamics suitable for use by non-mathematical major based researchers that even without writing down the system dynamics equations they can fill some forms and set some parameters and simulate and see the results. Therefore, we offer a three-step process for each system dynamics based on cell signaling modeling:

1. Problem Structuring (based on the equations for the reactions, kinetics rates and initial concentration of the reactants elements)
2. Causal loop diagram creation (for identifying the relations and the feedback loop structure)

3. Development of Dynamic Model including stock-flows and causal relations (for identifying quantities and generating dynamic simulation)

There are various software for system dynamics modeling such as: Dynamo, Powersim, Vensim, Stella, ithink, Anylogic, NetLogo and etc (Nuhoğlu & Nuhoğlu, 2007). The Models and outputs used in this paper are developed using Vensim system dynamics software PLE version[1], the free version.

### 2.3. Case Study: RKIP Influence on the ERK Signaling Pathway

There are plenty of suitable signaling pathways for investigation as a case study for this paper approach. We found RKIP Influence on the ERK Signaling Pathway a relatively small, easy to understand and well studied one which has been presented in several papers (Calder, Gilmore, & Hillston, 2006; Gilbert & Heiner, 2006; Kwang-Hyun et al., 2003).

The Ras/Raf-1/MEK/ERK pathway is a most important signaling pathway with crucial behavior in cell survival, proliferation and differentiation. Ras is stimulated by an external cause, then binds to and activates Raf-1 (to become "activated" Raf showing by Raf-1*) then it enables MEK and after that ERK. Raf-1 is a kind of Serin/Threonine kinase that phosphorylates and activates the MEK (MAP kinase kinase). The activated MEK, phosphorylates and activates ERK (extracellular signal-regulated kinase) (Lefloch, Pouysségur, & Lenormand, 2009; Martínez-Carpio & Trelles, 2010; Oda, Matsuoka, Funahashi, & Kitano, 2005; Zhang et al., 2011). ERK or MAPK (Mitogen Activated Protein Kinase) phosphorylate a variety of the proteins that leads to cell growth, survival and differentiation (Balan et al., 2006; Catalanotti et al., 2009; Steelman et al., 2011). Cell differentiation is controlled by this sequence of protein interactions, till the ERK is activated. The role of the kinase inhibitor protein RKIP in the behavior of this pathway is a field of experimental research: it is supposed that it inhibits activation of the RAF, so it can “dampen” down the ERK pathway. Certainly, there is much evidence that RKIP inhibits the malignant transformation by Ras and Raf oncogenes in cell cultures and it reduces the tumor in cancer. Thus, good models of these pathways are required to understand the role of RKIP and develop new therapies. Moreover, an understanding of the function and structure of this pathway may lead to more general results applicable to other pathways (Calder et al., 2006). A graphical representation of the pathway has been shown in Fig. 5.

[1] Vensim.com

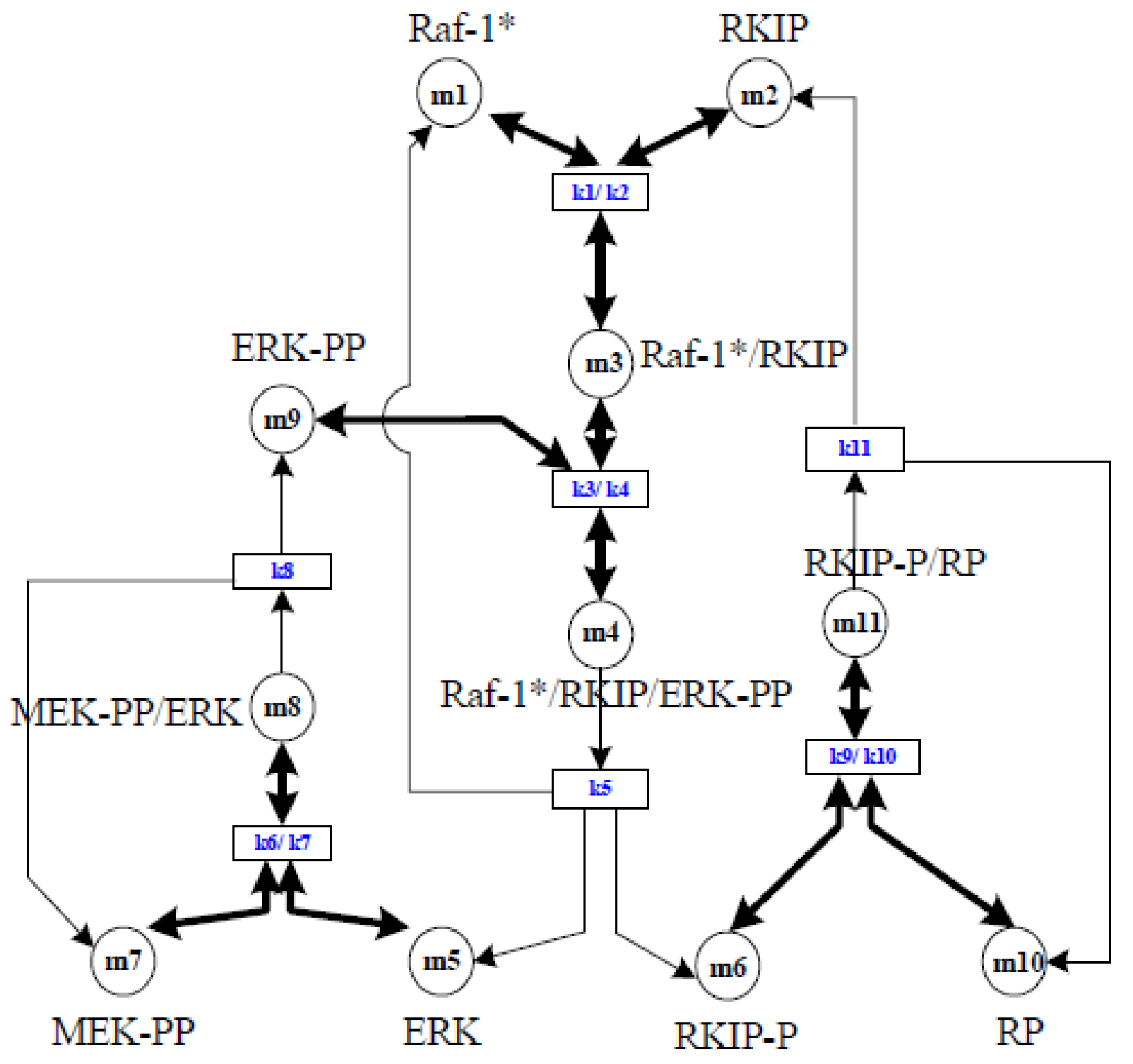

**Fig. 5.** Graphical representation of the ERK signaling pathway Regulation by RKIP (Kwang-Hyun et al., 2003)

As mentioned previously, there are three sets of parameters needed to completely simulate the model with the system dynamics approach, for a cell signaling pathway:

1. Equations for the reactions in the pathway (or equivalent mathematical differential equations)

2. Kinetics rate and constant parameters values
3. Initial concentration levels of the molecular species

We can construct the dynamic model directly by having 3 sets above (generally initial values for the stocks, parameter values and structural relationships in every model (Cavana & Maani, 2000)) unless we need more structural investigation, for example causal loop relations. In this case, is easily accessible by having the reactions in the pathway or the pathway graph. For Fig. 5, having the graphical representation of pathway equations, it is easy to sketch the causal loop diagram; Fig. 6. Positive or negative type of the feedback loops depends on the dynamic properties of the system, so it is not necessary to define the negative or positive property of the loop structure.

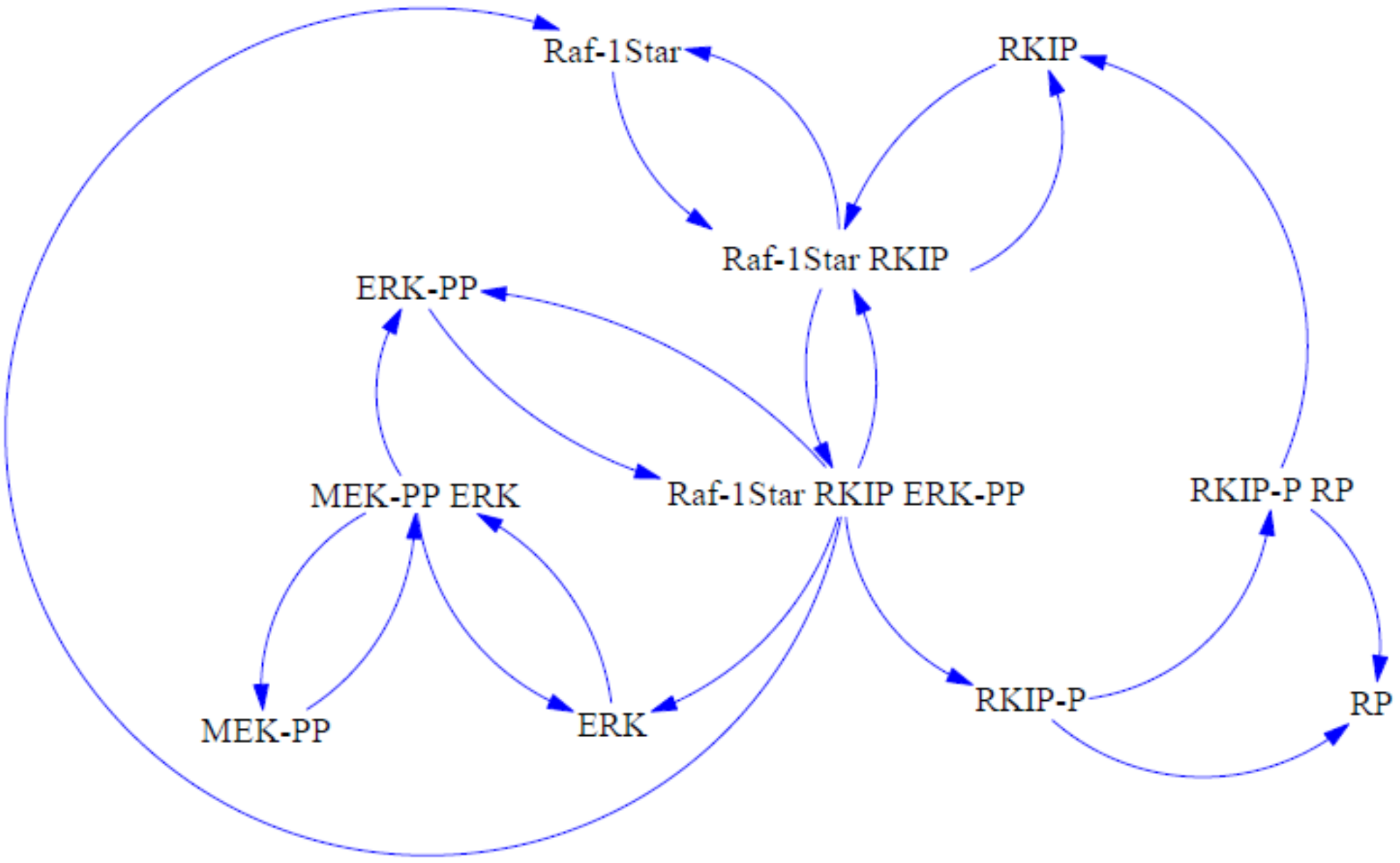

**Fig. 6.** Causal Loop Diagram for RKIP influence of ERK Signaling pathway

We use the followings to complete the model simulation (Tables 1 and 2) (Gilbert & Heiner, 2006). We select two different initial concentrations of species to compare two different runs of pathway with our modeling results, for more verification. Finally, the dynamic model can be seen in Fig. 7.

**Table 1.** Kinitics and the flows (rates) for the pathway (We use the abbrivations from Fig. 5 as synonyms for the lengthly species names) (Gilbert & Heiner, 2006)

| Parameters | Reaction Rates |
|---|---|
| $k_1 = 0.53$ | $r_1 = k_1 * m_1 * m_2$ |
| $k_2 = 0.0072$ | $r_2 = k_2 * m_3$ |
| $k_3 = 0.625$ | $r_3 = k_3 * m_3 * m_9$ |
| $k_4 = 0.00245$ | $r_4 = k4 * m4$ |
| $k_5 = 0.0315$ | $r_5 = k_5 * m_4$ |
| $k_6 = 0.6$ | $r_6 = k_6 * m_5 * m_7$ |
| $k_7 = 0.0075$ | $r_7 = k_7 * m_8$ |
| $k_8 = 0.071$ | $r_8 = k_8 * m_8$ |
| $k_9 = 0.92$ | $r_9 = k_9 * m_6 * m_{10}$ |
| $k_{10} = 0.00122$ | $r_{10} = k_{10} * m_{11}$ |
| $k_{11} = 0.87$ | $r_{11} = k_{11} * m_{11}$ |

**Table 2.** Initial two 'good' states of values for concentration of 11 species in the pathway (Gilbert & Heiner, 2006)

| **Species** | **S1** | **S2** |
|---|---|---|
| Raf-1* | 1 | 1 |
| RKIP | 1 | 1 |
| Raf-1*_RKIP | 0 | 0 |
| Raf-1*_RKIP_ERK-PP | 0 | 0 |
| ERK | 0 | 1 |
| RKIP-P | 0 | 0 |
| MEK-PP | 1 | 1 |
| MEK-PP_ERK | 0 | 0 |
| ERK-PP | 1 | 0 |
| RP | 1 | 1 |
| RKIP-P_RP | 0 | 0 |

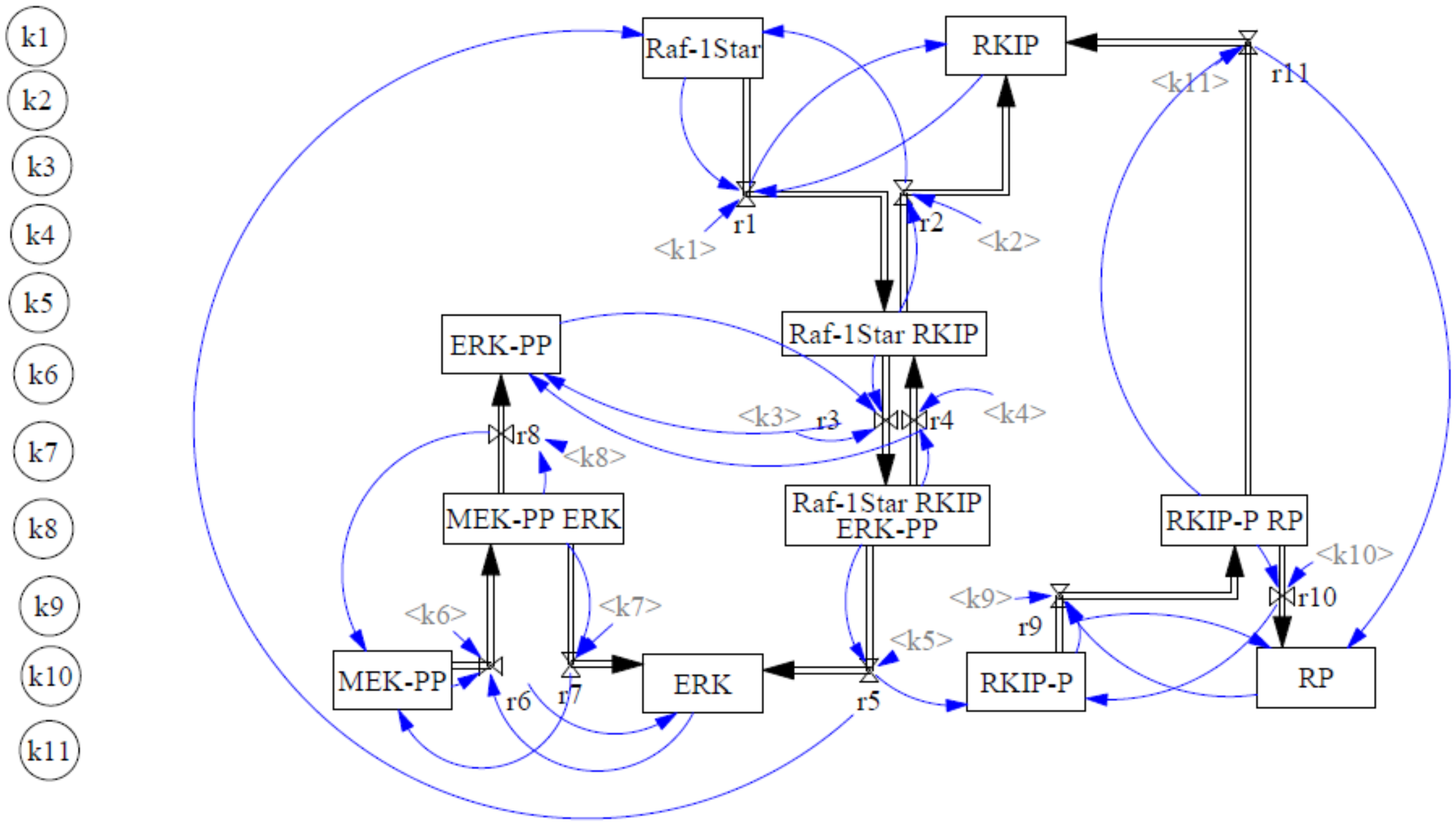

**Fig. 7.** Dynamic Modeling of ERK signaling pathway Regulation by RKIP

## 3. Results and Discussions

By making a causal loop diagram, we ought to obtain answers about the structure of the pathway, qualitative modeling. Modeling software can help us see *Causes Tree*, *Uses Tree* and *Loops*. *Causes Tree* creates a tree-type graphical representation showing the causes of workbench species. Reversely, *Uses Tree* creates a tree-type graphical representation showing the uses of a workbench species and *Loops* displays list of all feedback loops passing through a workbench species. For example, in case of an investigation of the effect of the RKIP on ERK, we can easily use *Causes Tree* to discover the causal influence paths between them in a tree structure (Fig. 8). This feature is especially useful in large complex pathways where it is not easy to discover the relations intuitively. Other modeling methods, such as Petri-Net, do not support such direct explicit mapping and clarity. One reason is that Petri-Net depends on user defined and model-imposed signs and symbols. In addition, *Loop* structure shows that “**Raf-1Star RKIP ERK-PP”** is shared between two loops consisting RKIP and ERK (Fig. 9). Thus, there is not a direct loop form RKIP to ERK.

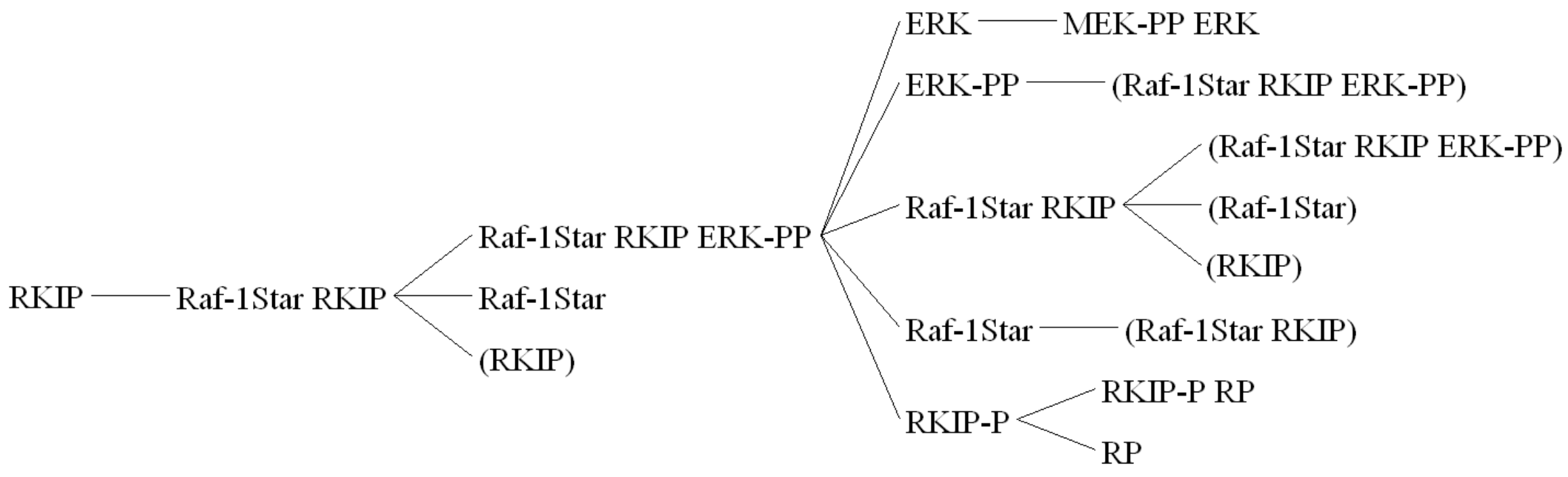

**Fig. 8.** Causes Tree revealing path between RKIP and ERK (Resulted by conection of two Causes Tree Diagram)

Loop Number 1 of length 1
Raf-1Star RKIP ERK-PP
Raf-1Star RKIP
Loop Number 2 of length 1
Raf-1Star RKIP ERK-PP
ERK-PP
Loop Number 3 of length 2
Raf-1Star RKIP ERK-PP
Raf-1Star
Raf-1Star RKIP
Loop Number 4 of length 3
Raf-1Star RKIP ERK-PP
ERK
MEK-PP ERK
ERK-PP
Loop Number 5 of length 4
Raf-1Star RKIP ERK-PP
RKIP-P
RKIP-P RP
RKIP
Raf-1Star RKIP

**Fig. 9.** Raf-1*_RKIP_ERK-PP Loops

Finally, for a more precise analysis of the pathway, dynamic behavior analysis, we compare our results with reference (Gilbert & Heiner, 2006), that has also modeled the influence of RKIP on ERK pathway with an integrative approach based on Petri Nets and ODEs, for two different runs of different initial concentrations (Table 2): S1 corresponding to the initial marking suggested by Cho et al. at (Kwang-Hyun et al., 2003) where the initial concentration of ERK-PP is high and ERK is low (Fig 10 to 11), and S2 corresponds to the initial marking suggested by (Gilbert & Heiner, 2006) with low ERK-PP and high ERK. Selecting only two states is for facilitating the comparison between our modeling and present modeling technique.

It can be concluded that the results are very similar, they are shown in Fig 13 to 15. Final states are equivalent for both initial concentrations. For more visible and comparable chart, we have divided our graphs into 2 parts, chart drawing and resolution. They are more clear and easy to follow when we use system dynamic software.

Simulation results in Fig. 12 show that ERK-PP (activated ERK) level decrease with sharp slopes in a way that after 20 seconds, ERK-PP has a noticeable drop and reaches 0.17, and MEK-PP (activated MEK) level decreases with a slow slope in the presence of RKIP. These results have high similarity with those given in (Gilbert & Heiner, 2006), Fig.10. For S2 state with the value of ERK-PP set to zero, considering the lack of ERK-PP, RKIP has no effect on it and ERK-PP level increases (Fig. 15). These results also coincide main reference, (Gilbert & Heiner, 2006). According to the simulation results depicted in Fig. 12 and Fig. 15, ERK-PP final concentration in both states is equal, lower than 0.17.

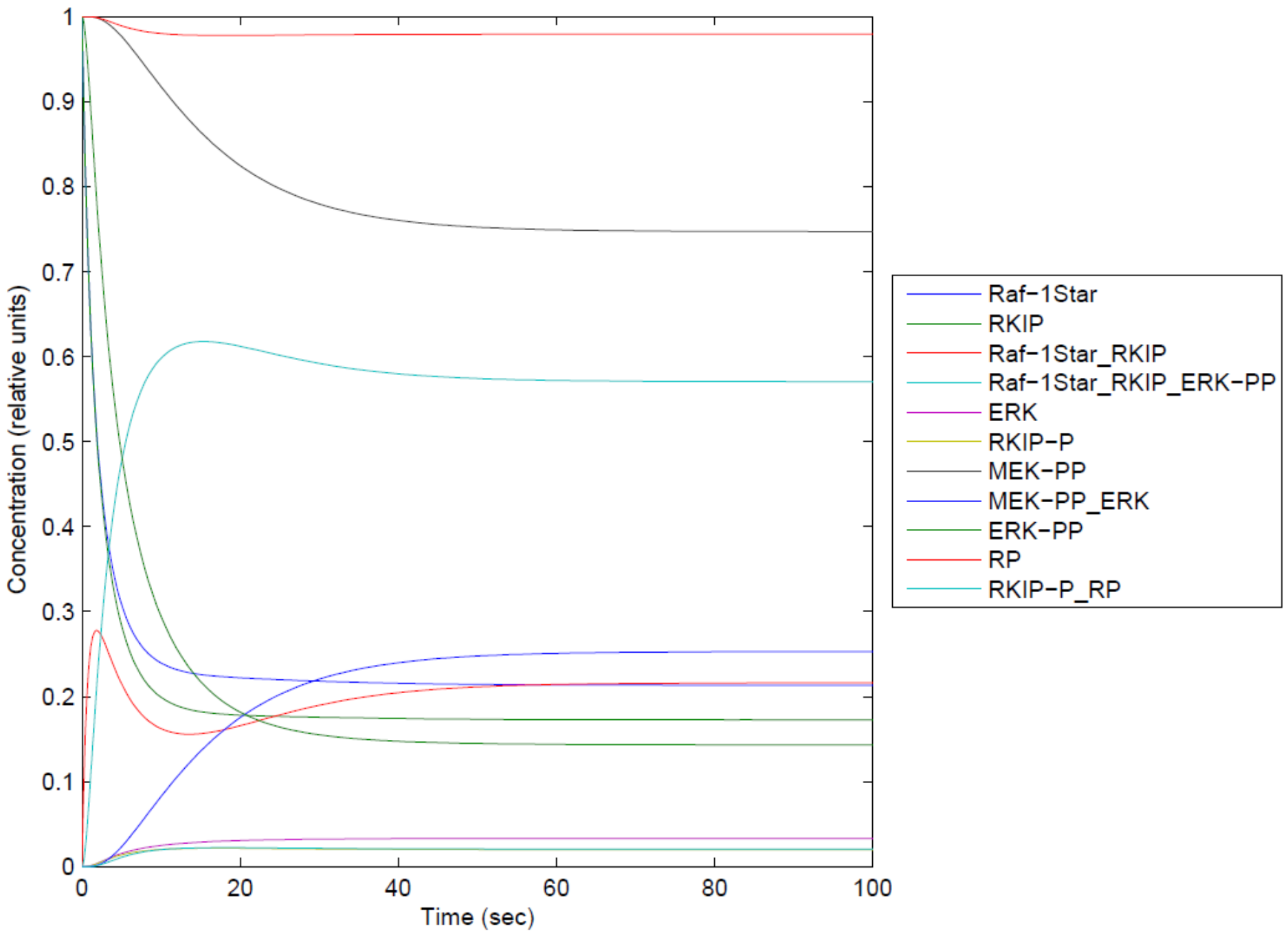

**Fig. 10.** Dynamic Behaviour of influence of RKIP on ERK pathway for State 1 where the initial concentration of ERK-PP is high and ERK is low (Gilbert & Heiner, 2006)

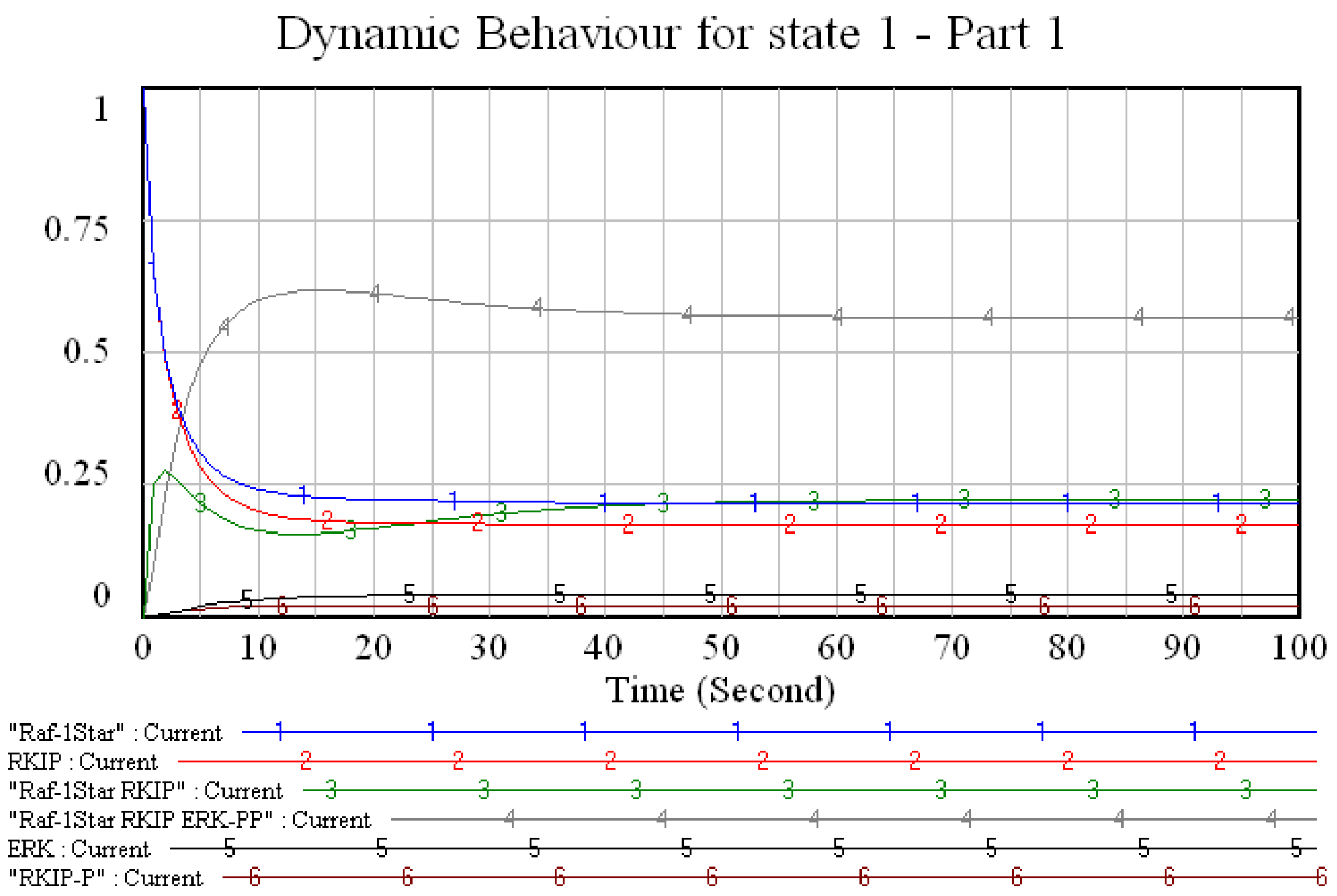

**Fig. 11** Dynamic Behaviour of influence of RKIP on ERK pathway for State 1 where the initial concentration of ERK-PP is high and ERK is low (Our simulation- part 1 of the output)

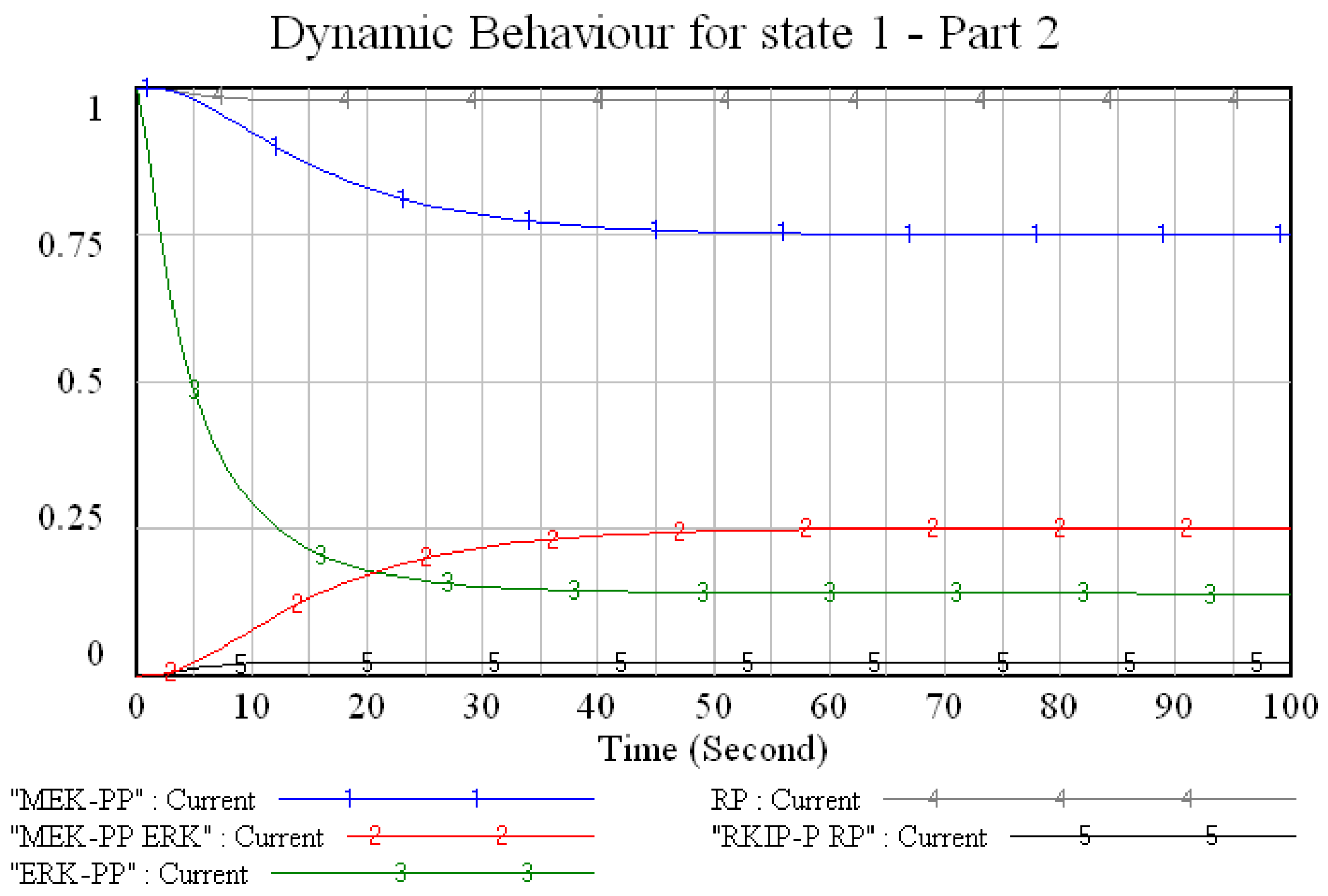

**Fig. 12** Dynamic Behaviour of influence of RKIP on ERK pathway for State 1 where the initial concentration of ERK-PP is high and ERK is low (Our simulation- part 2 of the output)

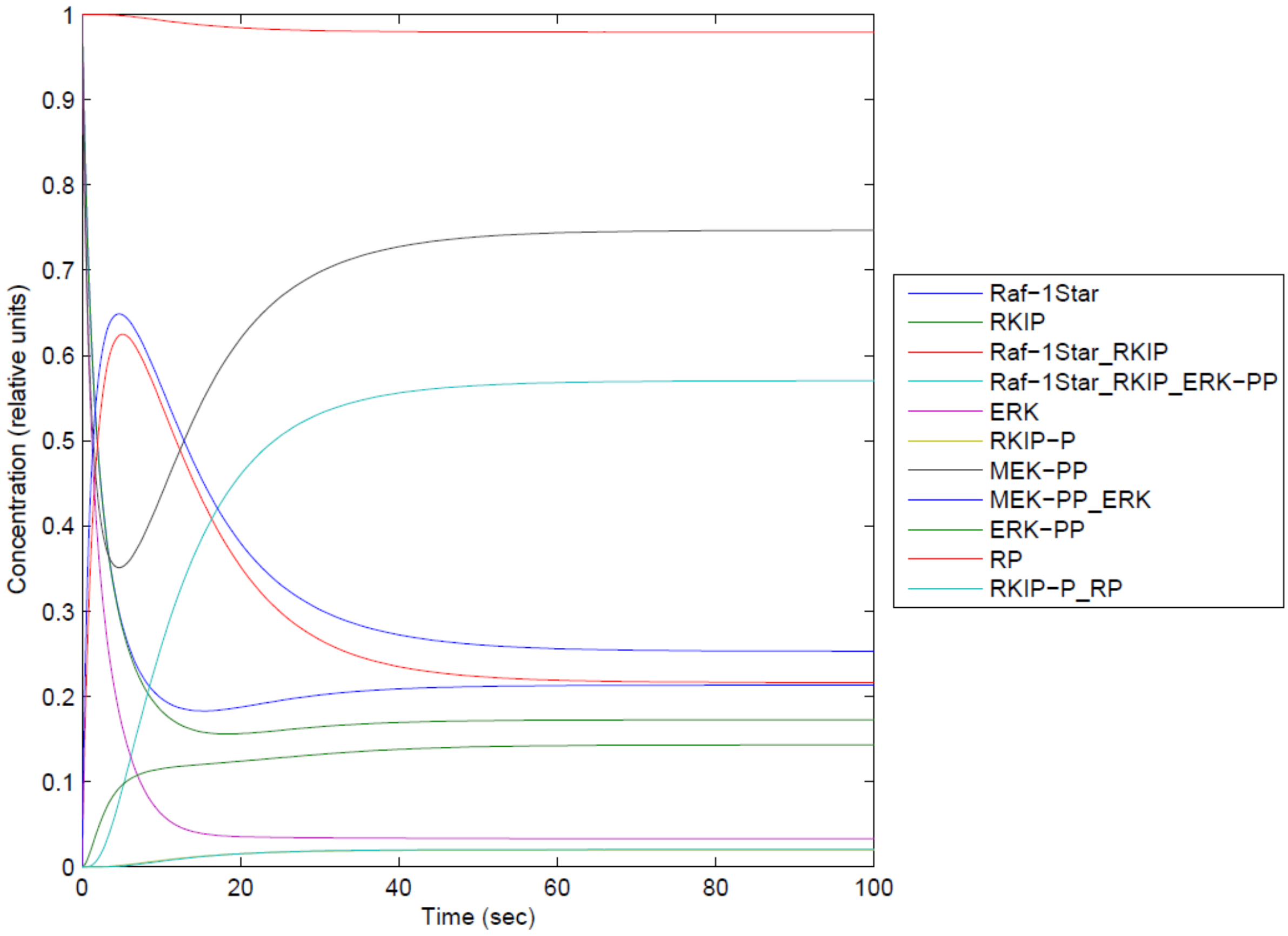

**Fig.** 13 Dynamic Behaviour of influence of RKIP on ERK pathway for State 2 where the initial concentration of ERK-PP is low and ERK is high (Gilbert & Heiner, 2006)

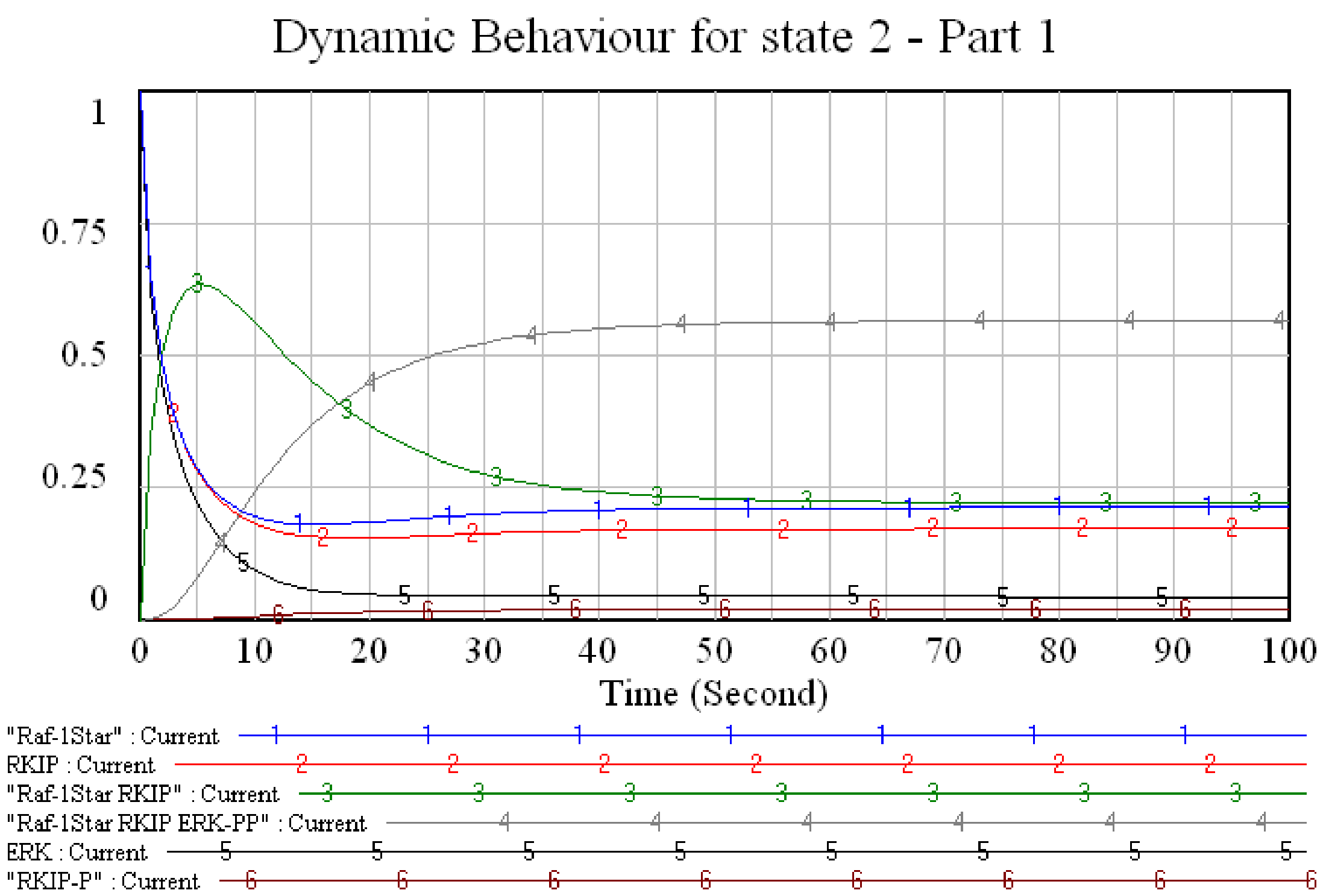

**Fig. 14** Dynamic Behaviour of influence of RKIP on ERK pathway for State 2 where the initial concentration of ERK-PP is low and ERK is high (Our simulation-part 1 of the output)

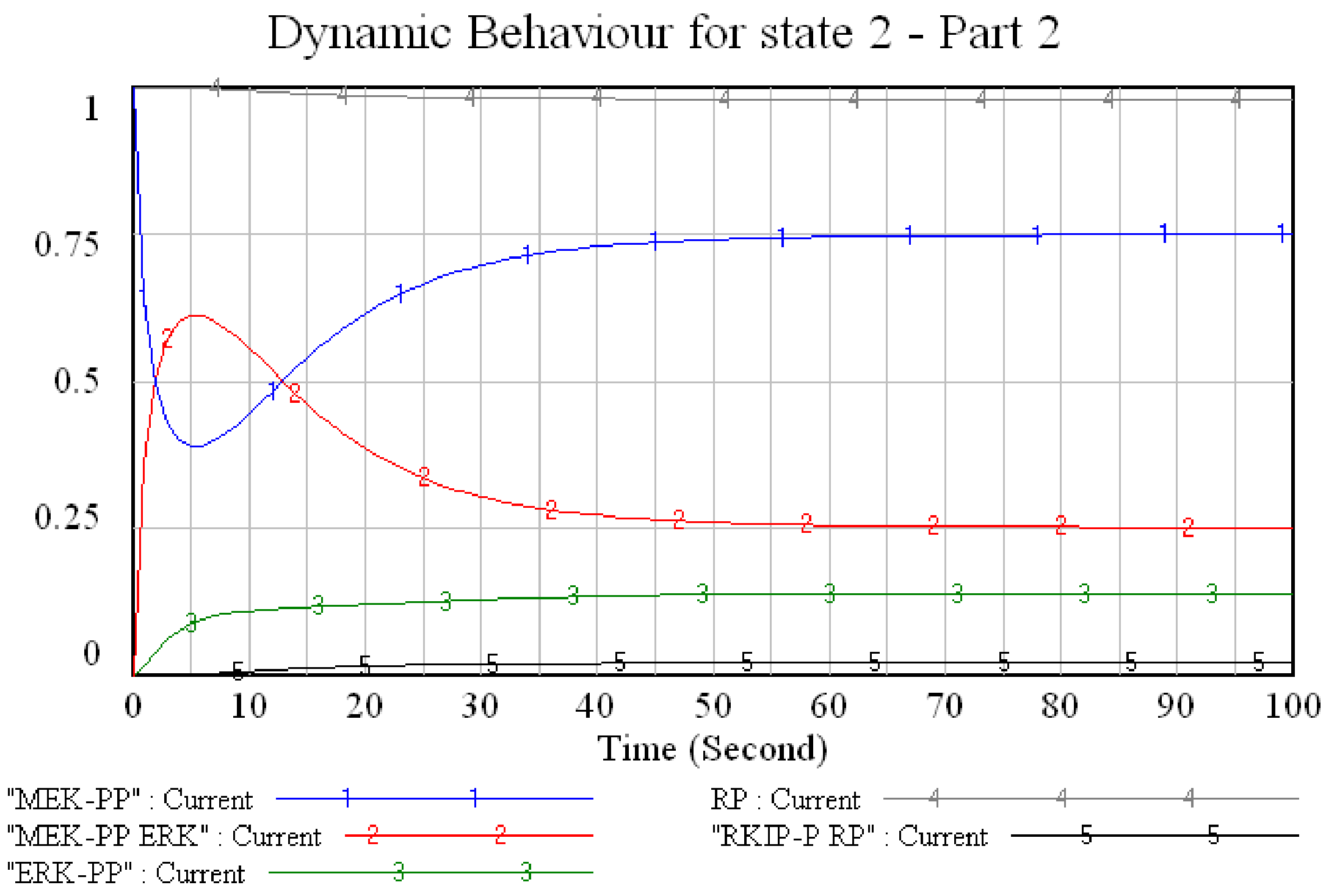

**Fig. 15** Dynamic Behaviour of influence of RKIP on ERK pathway for State 2 where the initial concentration of ERK-PP is low and ERK is high (Our simulation- part 2 of the output)

Fig. 15 shows the kinetics behavior effect of RKIP on MEK-PP for state 2. The dropping slope is sharper than Fig. 12, but the steady state concentration is the same, 0.75, for both figures.

Fig. 16 indicates RKIP binding to Raf-1* (activated Raf). It shows deactivation or decreasing levels of Raf1* with a steep slope. Raf-1* activity decrease, inactivation, finally leads to activity decrease, inactivation, of ERK, Fig. 12. Therefore, outputs show that the rate at which RKIP binds to Raf-1 Star affects the ERK pathway, as predicted (and observed). We can conclude that RKIP does indeed inhibit the ERK pathway as also shown by other studies (Calder et al., 2006; Kwang-Hyun et al., 2003; Popova-Zeugmann & Pelz, 2011). Using automatically simulated on change (SytheSim) option of the simulation software, it can be seen that if we increase the k2, RKIP will increase and its associative Raf-1 Star will also increase and as we expect ERK decreases. Also with the increase of k1 coefficient, RKIP and Raf-1 Star decreases and ERK increases. As a result, system dynamics modeling also confirms previous modeling results.

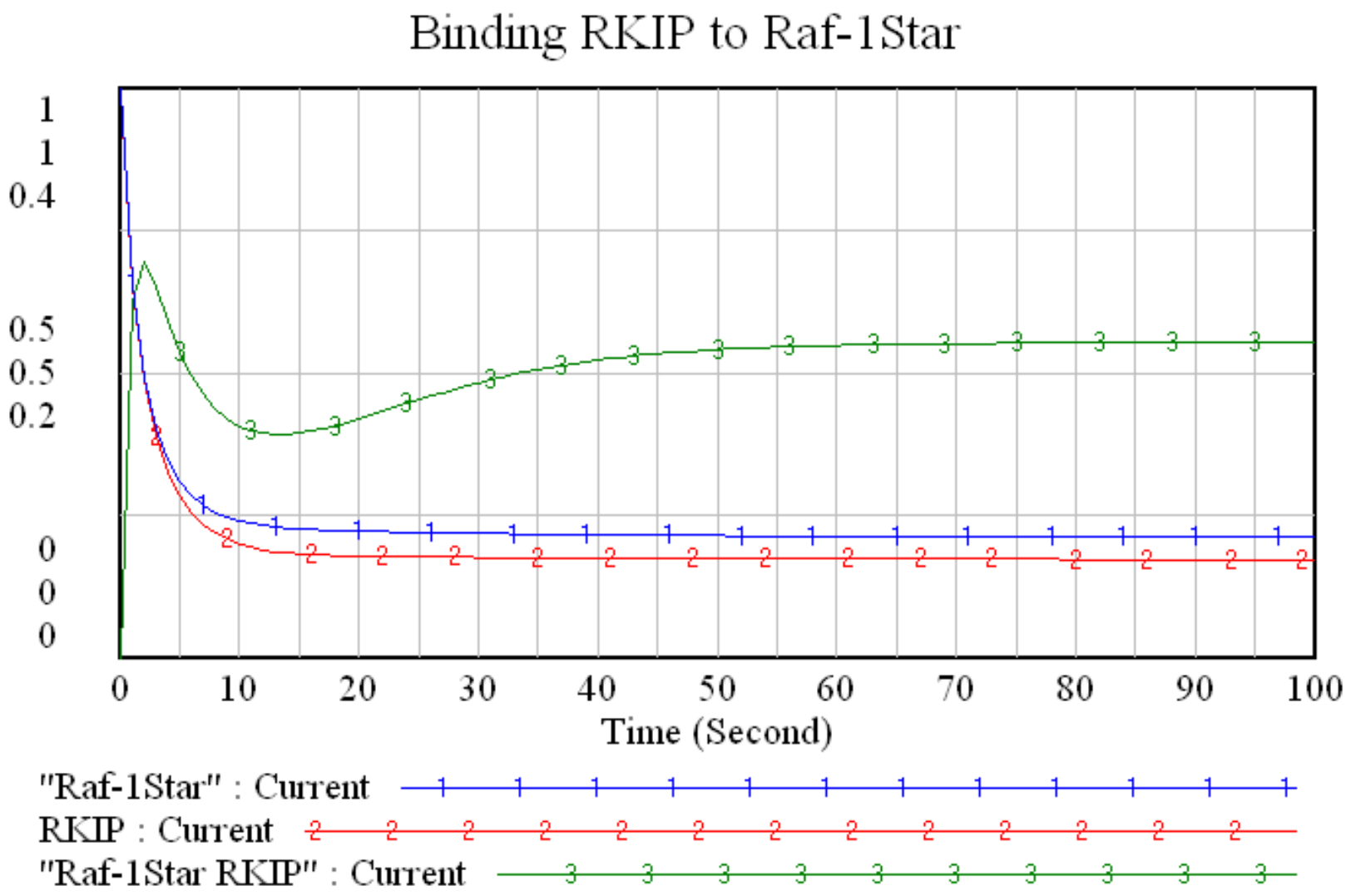

**Fig. 16** Dynamic behavior or RKIP Raf-1* Binding and their effects on ERK (S1 initial concentration)

### Conclusion

We applied for the first time, a well-known and widely-used dynamic modeling approach, system dynamics, successfully for analysis of a cell signaling pathway. We also described the process for more complex models. It has some considerable advantages including simplicity and effectiveness (Conference, 2006). For example, it can provide the researcher with useful insights into the system in order to understand it intuitively before quantitative analysis, which is achieved using causal loop diagrams and its derivatives. Having more investigation viewpoints also helps researchers to better understand the model. As such, they do not necessarily need to write down and solve related differential equations of the pathway; it is enough to draw and complete related charts in the system dynamics model. Results are produced in the form of graphs or numerical lists. Also, the parameter values of the model can be easily changed during simulation in order to get and compare new results and describe the dynamic behavior of the system. Furthermore, system dynamics can be used to verify another simulation model as an alternative modeling facility. There are, additionally options to use various functions, operations and types with system dynamics Stock-Flow elements that strength it to support more complicated states and features. Its associated qualitative and quantitative view of the system will give the researcher a comprehensive view of the pathway.

One limitation is requiring the kinetic rates for the simulating pathways in system dynamics modeling, while some signal transduction systems has defective information about the overall structure and molecular interactions. Without all the kinetics and detailed information about changes over the time, it is not easy to construct the model precisely (Albert & Wang, 2009). Another restriction is scalability. Our proposed model verified for a relatively small size example. For large signaling pathways with higher dimensional elements, we need to propose some modularization approach in order to reduce the complexity of the depiction and simulation. For example, we can gather relatively independent pathways in one package and behave it like an alone species in relation to other parts of the model in order to reduce the size of the investigated network. Therefore, using system dynamics universally for all the signal transduction networks needs more investigation, modification and justifications. It is proposed for the future works of this paper. The suggested technique will be used to simplify complex cell signaling pathways in systems biology. The simplified model is sometimes called the dominant system (DS). It gives the simplest possible model for the original model. The dominant system provides us the main asymptotic terms of the stationary state and relaxation time.

**Acknowledgements**

Many thanks to Hossein Toosi helped us with earlier version of the work.